\documentclass[a4paper,fleqn]{cas-dc} 
\usepackage[numbers,sort]{natbib}
\usepackage{hyperref}

\usepackage{caption}
\usepackage{subcaption}
\usepackage{lineno}

\def\tsc#1{\csdef{#1}{\textsc{\lowercase{#1}}\xspace}}
\tsc{WGM}
\tsc{QE}
\tsc{EP}
\tsc{PMS}
\tsc{BEC}
\tsc{DE}

\definecolor{antiquefuchsia}{rgb}{0.57, 0.36, 0.51}
 
\definecolor{bittersweet}{rgb}{1.0, 0.44, 0.37}

\definecolor{blue}{rgb}{0.1, 0.2, 1.00}

\begin{document}
\let\WriteBookmarks\relax
\def\floatpagepagefraction{0.9}
\def\textpagefraction{.001}

\title [mode = title]{DenResCov-19: A deep transfer learning network
for robust automatic classification of COVID-19, pneumonia, and tuberculosis from X-rays}
\shorttitle{DenResCov-19: A deep transfer learning network
for classification of CXRs}

%

\author[1,2]{Michail Mamalakis}[orcid=0000-0002-4276-4119]
\cormark[1]
\ead{mmamalakis1@sheffield.ac.uk}
\ead[url]{https://www.sheffield.ac.uk/dcs, https://insigneo.org/}
\shortauthors{M. Mamalakis et al.}
\address[1]{Department of Computer Science, University of Sheffield, Sheffield, UK}
\address[2]{Insigneo Institute for in-silico Medicine, Sheffield, UK}

\author[5,2]{Andrew J. Swift}

\author[6]{Bart Vorselaars}[orcid=0000-0001-5881-2341]

\author[7]{Surajit Ray}

\author[8]{Simonne Weeks}

\author[9]{Weiping Ding}

\author[1,2]{Richard H. Clayton}

\author[8]{Louise S. Mackenzie}

\author[3,4]{Abhirup Banerjee}[orcid=0000-0001-8198-5128]
\ead{abhirup.banerjee@cardiov.ox.ac.uk}
\ead[URL]{http://users.ox.ac.uk/~card0439/}
\cormark[1]

\address[3]{Division of Cardiovascular Medicine, Radcliffe Department of Medicine, University of Oxford, Oxford OX3 9DU, UK}
\address[4]{Institute of Biomedical Engineering, Department of Engineering Science, University of Oxford, Oxford OX3 7DQ, UK}
\address[5]{Department of Infection, Immunity \& Cardiovascular Disease, University of Sheffield, Sheffield, UK}
\address[6]{School of Mathematics and Physics, University of Lincoln, Brayford Pool, Lincoln LN6 7TS, UK}
\address[7]{School of Mathematics and Statistics, University of Glasgow, Glasgow G12 8QW, UK}
\address[8]{School of Pharmacy and Biomedical Sciences, University of Brighton, Brighton BN2 4GJ, UK}
\address[9]{School of Information Science and Technology, Nantong University, Nantong 226019, China}

\cortext[cor1]{Corresponding author}
%

\begin{abstract}
The global pandemic of COVID-19 is continuing to have a significant effect on the well-being of global population, increasing the demand for rapid testing, diagnosis, and treatment. Along with COVID-19, other etiologies of pneumonia and tuberculosis constitute additional challenges to the medical system. In this regard, the objective of this work is to develop a new deep transfer learning pipeline to diagnose patients with COVID-19, pneumonia, and tuberculosis, based on chest x-ray images. We observed in some instances DenseNet and Resnet have orthogonal performances. In our proposed model, we have created an extra layer with convolutional neural network blocks to combine these two models to establish superior performance over either model. The same strategy can be useful in other applications where two competing networks with complementary performance are observed. We have tested the performance of our proposed network on two-class (pneumonia vs healthy), three-class (including COVID-19), and four-class (including tuberculosis) classification problems. The proposed network has been able to successfully classify these lung diseases in all four datasets and has provided significant improvement over the benchmark networks of DenseNet, ResNet, and Inception-V3. These novel findings can deliver a state-of-the-art pre-screening fast-track decision network to detect COVID-19 and other lung pathologies.



\begin{highlights}
\item A pre-screening fast-track decision network to detect COVID-19 and other lung pathologies.
\item A new deep-learning network, named DenResCov-19, for robust and accurate classification.
\item Evaluating the accuracy and robustness of DenResCov-19 over heterogeneous chest x-ray image datasets with binary and multi-class labels (COVID-19, pneumonia, tuberculosis, and healthy).
\item Evaluating the robustness of DenResCov-19 over a Monte Carlo cross validation scheme for multi-class classification.
\item Comparison of DenResCov-19 with established networks of ResNet-50, DenseNet-121, VGG-16, and Inception-V3. 

\end{highlights}


\maketitle

\section{Introduction}
Coronavirus 2019 (COVID-19), a disease caused by severe acute respiratory syndrome coronavirus 2 (SARS-CoV-2) virus, has affected the health of populations globally \cite{1}. In order to control the COVID-19 pandemic, there is an urgent need for rapid and accurate diagnostic testing in healthcare \cite{3,2}. Since SARS-CoV-2 can cause COVID-19 pneumonia and severe lung damage, differentiating viral from bacterial pneumonia and other respiratory infections such as Tuberculosis (TB) using chest imaging technology is essential for managing infection control decisions and diagnosis and for planning treatment regimes \cite{4}.

Many infectious respiratory diseases present in a similar manner, with symptoms such as difficulty in breathing, persistent cough, and fever. Pneumonia, an infection affecting the airspaces in the lung, is caused by various etiological agents such as bacteria, viruses, and fungi. There is wide range of symptoms associated with the infection, which include shortness of breath, fever, phlegm production, and cough. The progression of the disease is marked by the air space opacification, which can be detected using imaging diagnostics \cite{5,6}. Despite the availability of antimicrobials, pneumonias contribute to the most common cause of mortality, especially in childhood \cite{7}. In addition, TB induces a persistent cough and breathlessness with symptoms that overlap those of pneumonia and COVID-19. The mortality rates have also risen due to drug-resistant pulmonary TB, caused by Mycobacterium tuberculosis \cite{8}. There is therefore a clear need for a robust artificial intelligence (AI) system that can detect and classify the various respiratory diseases that have overlapping presentations to the clinic, so that the right course of treatment regime can be prescribed.

The standard imaging modalities for lung disease diagnosis include magnetic resonance imaging (MRI), chest X-ray (CXR), and computed tomography (CT) scan. Although MRI and CT scan are the gold standard for assessing lung diseases, they are more expensive, involve radiation exposure, and not readily available globally \cite{9}. In comparison, CXR is less expensive, readily available, and  is one of the most common diagnostic imaging techniques for cardiothoracic and pulmonary disorders.

CXR patterns of lung disease present differentiation challenges and often result in high inter-reader variability across radiologists \cite{10}. With potential future waves of the pandemic, radiologists' workloads will increase and there is an urgent need for new automated image analysis tools that can enhance the radiologists' qualitative assessment. These tools will classify or segment sections of the CXR in order to support the diagnostic workflow. Decision support systems are designed to aid the clinical decision-making process and have established themselves as emerging research trend in healthcare \cite{11}. Over recent months of the pandemic, automated detection of pneumonia or other lung diseases, specifically their early detection and classification, have gained significant attention from both clinical and the AI researchers.

The development of AI-based medical systems, as well as their translation to medical practice, is playing an increasingly prominent role in the treatment and therapy of patients \cite{GREENSPAN2020}. Along with the automated methods that rely on the blood test results or biomarkers for diagnosis \cite{BANERJEE2020,SONG2020,LAL}, an increasing number of deep learning-based methods, specifically the convolution neural network (CNN)-based models \cite{cov2,Das2020,densenet_idea,covCT2,OZTURK2020}, are being implemented and used in order to develop accurate, robust, and fast detection techniques to fight against COVID-19 and other respiratory diseases.

In this regard, the aim of the current study is to test the feasibility of early automated detection and distinction between COVID-19, pneumonia, TB, and healthy patients based on CXR scans. We have developed a deep transfer learning pipeline, named DenResCov-19, to diagnose if a patient is healthy or has a lung disease. The proposed network optimally combines the DenseNet-121 and ResNet-50 networks. This combination unifies the simplicity of ResNet structure and the complexity of DenseNet blocks and delivers a well-balanced result of accuracy and increased specificity and sensitivity. Pretrained networks on the ImageNet cohort are used as transfer learning techniques. We have tested the adaptability of our proposed network for two-class (pneumonia and healthy), three-class (COVID-19 positive, pneumonia, and healthy), as well as four-class (COVID-19 positive, pneumonia, TB, and healthy) classification problems. To the best of our knowledge, this is the first work to examine the feasibility of early automatic detection and distinction between COVID-19 positive, pneumonia, TB, and healthy patients based only on CXR scans using a DL network. The proposed DenResCov-19 network has been able to perform optimally in different multi-class problems and has achieved robust and improved performance over the state-of-the-art methods for the classification of lung-diseases in all our datasets. The main contributions of this paper are as follows:
\begin{enumerate}
\item The development of a new deep-learning network, named DenResCov-19, for robust and accurate classification.
\item Evaluating the accuracy and robustness of DenResCov-19 over heterogeneous CXR image datasets with binary and multi-class labels (COVID-19, pneumonia, TB, and healthy).
\item Evaluating the robustness of DenResCov-19 network over a Monte Carlo cross validation scheme for multi-class classification.
\item The comparison of DenResCov-19 with established networks of ResNet-50, DenseNet-121, VGG-16, and Inception-V3. 
\item Developing a pre-screening fast-track decision network to detect COVID-19 and other lung pathologies.
\end{enumerate}

\par The rest of the paper is organised as follows: Section~\ref{sec:related-work} gives a brief overview of the related works. Section~\ref{sec:methodology} presents the development of the proposed methodology, while Section~\ref{sec:implementation} summarises its implementation and the description of clinical datasets. Numerical results of the application of our method on four different datasets are presented in Section~\ref{sec:results}. The final conclusions are presented in Section~\ref{sec:conclusions}.

\section{Related Works}
\label{sec:related-work}
In this section, we present a brief overview of pneumonia and COVID-19 diagnosis studies, based on CXR and CT scans.

\subsection {Review of pneumonia detection in CXR images}
There exists a significant body of literature on the application of deep learning networks on CXR images for detecting pneumonia in patients \cite{BUSTOS2020,JAISWAL2019,Varshni2020,ISJ}. Here we give a summary of the most important approaches.

Jaiswal et~al. \cite{JAISWAL2019} used Mask-Region-based CNN \cite{mcnn} model to automatically identify potential pneumonia cases from CXR images. Bharati et~al. \cite{pneumonia} proposed a hybrid deep learning framework by combining VGG \cite{vgg}, data augmentation, and spatial transformer network (STN) with CNN. They trained their model in NIH CXR dataset \cite{KERMANY2018} with $73\%$ accuracy. Even though their approach did not achieve a high accuracy, their network required training time of only $431$~s on their full dataset. Bustos et~al. \cite{BUSTOS2020} presented a comprehensive study on a significantly large dataset of $160,000$ CXR images, including $19$ different classes of lung diseases. They compared four models, namely CNN, recurrent neural network (RNN) composed of bi-directional long short-term memory (LSTM) cells \cite{ho}, CNN with per-label attention mechanism (CNN-ATT) \cite{catt}, and RNN composed of bi-directional LSTM cells with per-label attention mechanism (RNN-ATT). Among the four models, the RNN-ATT model achieved the best results with $86.4\%$ accuracy with only $41$ epochs training. Varela-Santos and Melin \cite{ISJ} implemented an automated system for future detection of COVID-19 disease in CXR and CT lung images. They efficiently utilised the image texture feature descriptors from CXR images in feed-forward and convolutional neural networks for detecting COVID-19 and healthy individuals.

\subsection {Review of COVID-19 detection in CXR and CT images}
Prior to COVID-19, deep learning (DL) models have been used extensively for the classification of pneumonia and other lung  diseases. Following their successes, a range of DL approaches have been  developed  for diagnosing and differentiating COVID-19 lung infections \cite{he2021,GILANIE2021}. Most of these new approaches are based on CXR and CT modality, which are the most widely used imaging modality for diagnosing pneumonia and COVID-19 \cite{Das2020,densenet_idea,covCT2}. Here we review the performance of some of these studies.

\par Ozturk et~al. \cite{OZTURK2020} proposed the DarkCovidNet model to assist clinicians and radiologists to diagnose COVID-19. Their network, inspired by DarkNet, achieved accuracy of $98.08\%$ and $87.02\%$ respectively, for binary (COVID-19 vs healthy) and multi-class classification (COVID-19, pneumonia or healthy). DarkCovidNet is based on the DarkNet, which is good for fast performance (e.g., with self-driving cars); but in our case, time is not really a critical issue. Moreover, the network was only tested on a limited number of cases. Larger datasets will be able to test its robustness.  
 
\par Pereira et~al. \cite{cov2} designed the network model RYDLS-20, that achieved F1 value of $89\%$ for COVID-19 diagnosis. Their dataset was highly imbalanced, with $1000$ healthy cases and $90$ patients affected by COVID-19. More importantly, their classification performance was presented without any cross-validation step.

\par Yoo et~al. \cite{cov1} proposed a combination of three decision-tree classifiers for pre-screening fast-track decision making in order to detect COVID-19. Their pipeline was a combination of three binary decision trees, each trained by a deep learning model with CNN. The accuracies of the binary decision trees ranged between $80\%$ to $98\%$. However, their network did not test any pathologically confirmed data. In addition, they did not incorporate any data augmentation technique during training in order to reduce the overfitting effects. A large dataset of $5000$ CXR scans was used by Minaee et~al. \cite{Minaee2020} for classification of healthy and COVID-19 cases. They used four different models, including ResNet18 \cite{khe}, ResNet50 \cite{khe}, SqueezeNet \cite{sq}, and DenseNet-161 \cite{De}, and achieved on average a sensitivity of $98\%$ and specificity of $92\%$.

\par Another set of studies have presented satisfactory outcomes in the classification of COVID-19 and healthy cases from CT images \cite{chen2020,covCT1,cov5,covCT3,PPR:PPR113467}. Li et~al. \cite{cov5} studied CT images deployed at sixteen different hospitals. They used a U-net to first segment the lung regions and then applied a ResNet-50 to classify the patient as COVID-19 affected or not. Their pipeline achieved a good accuracy, because there was no noise in peripheral organ regions due to the segmentation of lungs. One of the limitations of their study is that their dataset had higher number of positive cases that made the prediction biased ($723/1136$). However, the major achievements of this study are the incorporation of inter-hospital variations in datasets and the use of six independent experts to arrive at the ground truth.

\par In another study, Li et~al. \cite{covCT2} used statistical methods, which included `total severity score' to classify healthy and unhealthy patients based on CXR images. The authors applied the Wilcoxon-rank test to predict the level of severity of the patients. They computed their ground truth using inter-scan and inter-observer variability and also provided thorough details on how the severity level was computed. However, their severity dataset was not large enough and also, they did not incorporate any data splitting based on advanced age, underlying diseases, and pleural effusions.

\par Song et~al. \cite{covCT3} implemented a deep learning-based CT diagnosis system, named Deep-Pneumonia, to identify patients with COVID-19. They manually segmented the lung region and then classified COVID-19 or healthy cases using a DL network. This network, named DRE-Net, is a combination of ResNet-50, feature pyramid network (FPN), and an attention module. The main advantages of this study are the multi-vendor datasets from three different hospitals, the very high sensitivity ($95\%$) and specificity ($96\%$) values, and the fast diagnosis time per patient ($30$~s). However, its drawbacks include: the need of semi-automatic lung segmentation, the classification of datasets based only on CT images without any splitting depending on advanced age, underlying diseases, or pleural effusions, and the absence of any reference in inter-observer variability of the ground truth.

\par Chen et~al. \cite{chen2020} trained their deep network using $46,096$ anonymous images from $106$ admitted patients, including $51$ patients with laboratory confirmed COVID-19 pneumonia and $55$ control patients with other diseases, in Renmin Hospital of Wuhan University. They used a U-net++ network to segment the lungs and classified whether the region had a scar area. The statistical comparison of two-tailed paired Student's $t$-test with $0.05$ significance level was used for time comparison between radiologist and the model. The key advantages of this study include: the large and well-balanced training dataset, the high classification accuracy (over $95\%$), and the use of three expert radiologists accounting for the inter-observer variability to extract the ground truth. The main limitations of this study were that the dataset was collected from only one hospital and the classification was based only on CT images without any data-splitting based on advanced age, underlying diseases, or pleural effusions. Also, their lung segmentation step, being in a new cohort, can potentially decrease the total classification accuracy.

\par As discussed, some studies have attempted to solve the problem of automated diagnosis of pneumonia and COVID-19, based on existing deep learning networks \cite{Minaee2020,OZTURK2020,cov2} on CT  \cite{chen2020,covCT1,cov5,covCT3,PPR:PPR113467} or on CXR \cite{OZTURK2020,cov1,Das2020,covCT2,densenet_idea} image cohorts. However, the noted algorithms suffer from the following limitations and challenges:
\begin{enumerate}
\item The lack of regularization techniques (data augmentation, penalty norms, etc.) used in models to avoid possible overfitting.
\item Lack of balance in the models between the speed and the robustness and accuracy.
\item The lack of generalization  techniques, such as cross-validation, for accurate prediction of the models.
\item The need of manual segmentation of the lung region from experts to deliver a robust semi-automatic classification result.
\item The validation of the models for only binary classification \cite{Minaee2020,cov2,cov1} or three-class (COVID-19, pneumonia, and healthy) classification tasks \cite{OZTURK2020,Das2020,covCT2,densenet_idea}. 
\item The validation of the models in one specific cohort (\textit{i.e.} no cross-vendor or cross-institute validation).
\end{enumerate}

\begin{figure*}
 \centerline{\includegraphics[width=0.99\textwidth]{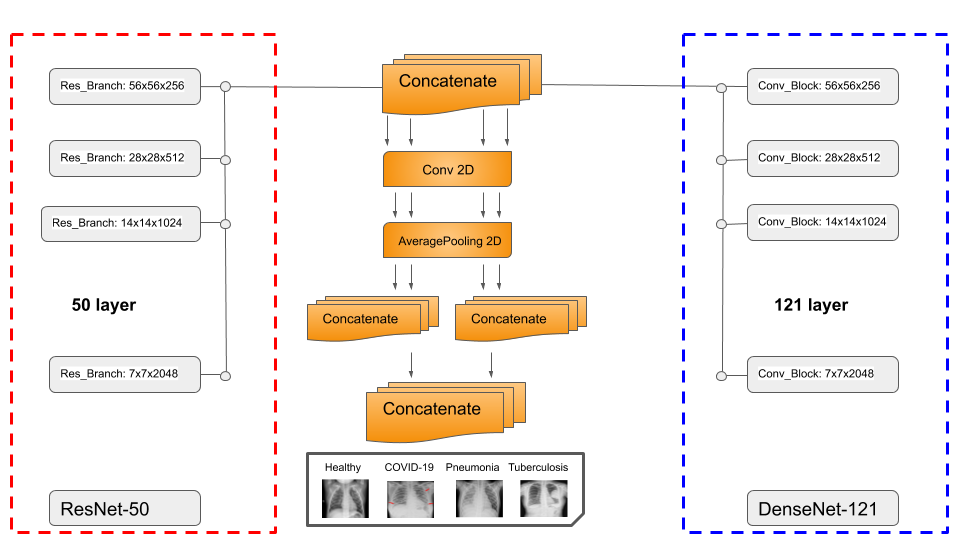}}
 \caption{DenResCov-19: a deep transfer learning pipeline to classify if a patient has COVID-19, pneumonia, or tuberculosis, based on CXR.}
 \label{pipre}
\end{figure*}

\section{Methodology and Background}
\label{sec:methodology}
In the current study, we propose to  train a deep learning network, named DenResCov-19, to solve a multi-class problem, namely,  whether a patient is healthy or has pneumonia, COVID-19, or tuberculosis. 

\subsection{Background}
Our approach is based on two state-of-the-art networks: ResNet \cite{he} and DenseNet \cite{De}. They have recently been used to solve similar multi-class problems.

\par \textbf{ResNet-$L$} is inspired by the structure of VGG nets \cite{VGG2}. The network comprises of $L$ layers, each of which implements a non-linear transformation. In the majority of ResNet-based networks, the convolutional layers have $3 \times 3$ filters. Downsampling is performed by convolutional layers with a stride of $2$. The last two layers of the network are an average pooling layer, followed by a $1000$-way fully-connected (FC) layer. The main rule of this deep network is that the layers have the same number of filters as the number of the output feature map size. In case the feature map size is halved, the number of filters is doubled, thus reducing the time complexity per layer.
CNN feed-forward inputs \(x_{i}\) are the outputs \(x_{i-1}\) of the previous layer, so the transition layer is given by $x_i = H_i(x_{i-1})$. In particular, ResNet adds a skip-connection and the identity function is given by:
\begin{equation}
x_i = H_i(x_{i-1}) + x_{i-1}
\label{1}
\end{equation}

\par \textbf{Densenet-$L$} is a convolutional network. The network comprises of $L$ layers, each of which implements a non-linear transformation. These transformations can be different function operations, such as Batch Normalization, rectified linear units (ReLU), Pooling, and Convolution. Huang et~al. \cite{De} introduced a unique connectivity pattern information flow between layers to direct connecting any layer to all subsequent layers. As a result, the $i$th layer includes the feature-maps of all previous layers. The input of $i$th layer is given by the equation:
\begin{equation}
x_i = H_i\left([x_0, x_1,\cdots, x_{i-1}]\right)
\label{2}
\end{equation}
where $[x_0,x_1,\cdots, x_{i-1}]$ refers to the concatenation of the feature-maps produced in layers $0,\ldots,i-1$.  All inputs of a composite function $H_i(\cdot)$ are concatenated into a single tensor. Each composite function is a combination of batch normalization (BN), followed by a rectified linear unit (ReLU) and a $3 \times 3$ convolution (Conv).

\subsection{Network architecture}
To evaluate the state-of-the-art networks before we train and test them in the CXR cohorts, we initially test them in the open-source and widely used CT cohort of \cite{CT}. Since there is currently a lack of existing publicly available dataset of CXR images relating to COVID-19 cases, we have tested the behavior of benchmark models in the CT cohort in order to check if the expected behavior of the proposed network can be observed (i.e. achieve high F1 and AUC-ROC values).

\begin{table}
\caption{Metrics of deep learning networks to classify pneumonia, COVID-19, and healthy cases in CT images}
\centering
\begin{tabular}{ |p{2.05cm}||p{1.95cm}|p{1.45cm}|p{1.2cm}| }
    \toprule
 \multicolumn{4}{|c|}{CT dataset \cite{CT} } \\
\cmidrule(r){1-4}
 Metric (\%) & DenseNet-121 & ResNet-50 & VGG-16 \\
 \midrule
Recall      & 44.0 & 71.2 & 100.0 \\
Precision   & 81.2 & 91.0 & 50.0\\
AUC-ROC     & 86.4 & 64.0 & 51.0 \\
F1          & 58.4 & 81.0 & 71.4 \\
\bottomrule
\end{tabular}
\label{t1}
\end{table}

Table~\ref{t1} highlights the results of DenseNet-121, ResNet-50, and VGG-16 networks for classification of pneumonia, COVID-19, and healthy cases in CT images. From the results, it can be observed that, while the ResNet has better recall, precision, and F1 metrics than the Densenet and VGG, Densenet has better AUC-ROC. Based on these observations, we hypothesize that a combination of the two models can deliver a well-balanced AUC and F1 metric results.

\begin{figure*}
 \centering{\includegraphics[width=.8\textwidth]{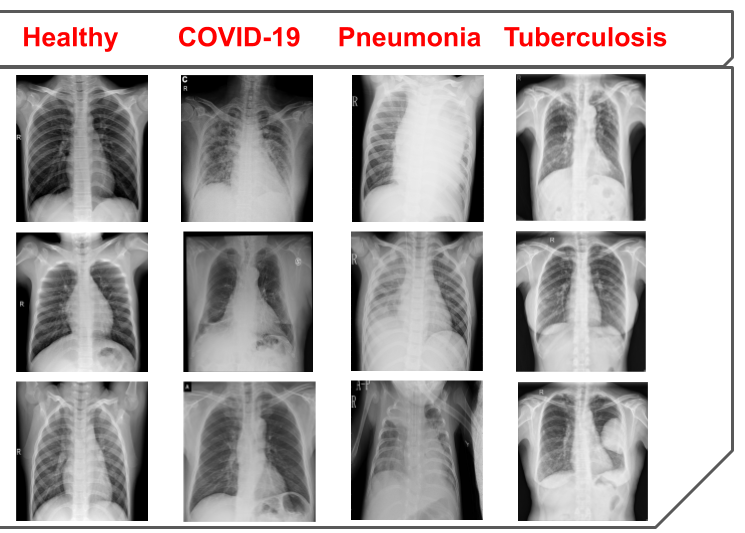}}
 \caption{A sample of healthy, COVID-19, pneumonia, and tuberculosis cases from the CXR image dataset.}
 \label{cxr}
\end{figure*}

The architecture of our proposed DenResCov-19 network is presented in Fig.~\ref{pipre}. DenResCov-19 network is a concatenation of four blocks from ResNet-50 and DenseNet-121 with width, height, and frames of $58 \times 58 \times 256$, $28 \times 28 \times 512$, $14 \times 14 \times 1024$, and $7 \times 7 \times 2048$, respectively. We chose these specific blocks from the networks, as we needed layers with the same width $\times$ height $\times$ frames, so that the information of both models can be combined. As a result, we used four different layers of 58, 28, 14, and 7 size kernels, as we wanted to concatenate the information of the two networks in different regions of interest. Each of the four outputs feed a block of convolution and average pooling layers. Thus, the initial concatenate information can be translated into the convolution space. After that, we used some levels of concatenation-CNN block techniques to create kernels that will deliver a final layer of soft-max regression, so that the network can conclude in the classification decision.

The convolution layer is defined as:
\begin{equation}
x_{i,j}^l = \sum_{a=0}^{M-1}\sum_{b=0}^{M-1} \omega_{ab} y_{(i+a)(j+b)}^{l-1},
\label{5}
\end{equation}
where $x_{i,j}^l$ is a unit in layer $l$, $\omega_{ab}$ is a $M \times M$ filter, and $y_{(i+a)(j+b)}^{l-1}$ is the nonlinearity of previous convolutional layer given by:
\begin{equation}
 y_{ij}^{l} = \sigma(x_{ij}^l).
\label{6}
\end{equation}

\par The average pooling layer is defined over a $K \times K$ region and outputs a single value, which is the average over that region. The inputs of the $l$th ($l=1, 2, 3, 4$) layer block are provided according to the equation:
\begin{equation}
x_l = H^{des}_l\left([x^{des}_0, x^{des}_1,\cdots, x^{des}_{i-1}]\right) + H_l^{res}\left(x^{res}_{i-1}\right) + x^{res}_{i-1},
\label{3}
\end{equation}
where $H^{res}_l(\cdot)$ is the composite function of $l$th  ResNet layer and $H^{des}_l(\cdot)$ is composite function of $l$th DenseNet layer. The last step of the pipeline is the combination of two pair concatenation and a global concatenation followed by a 512-way fully-connected softmax layer.

\subsection{Evaluation metrics}
The most common metrics for evaluating classification performance are the precision, recall, and F1-Score, which follow the standard definitions:
\begin{eqnarray}
\mbox{Precision}=\frac{TP}{TP+FP},\\
\mbox{Recall}/\mbox{True Positive Rate}=\frac{TP}{TP+FN},\\
\mbox{False Positive Rate} = \frac{FP}{FP+TN},
\end{eqnarray}
where $TP$, $TN$, $FP$, and $FN$ are the true positive, true negative, false positive, and false negative values, respectively. The F1-score is defined as the harmonic mean of the precision and recall, as follows
\begin{equation}
F1 = \frac{2 \times \mbox{Precision}\times \mbox{Recall}}{\mbox{Precision}+\mbox{Recall}} = \frac{2 TP}{2 TP + FP +FN}.
\end{equation}
Besides these metrics, we have also used the AUC-ROC metric values \cite{roc} for evaluation. The AUC (area under the curve)-ROC value can be computed by integrating over the receiver operating characteristic (ROC) curve, plotting the true positive rate against the false positive rate.

\section{Implementation}
\label{sec:implementation}
This section describes the implementation details of the proposed DenResCov-19 pipeline.

\subsection{Cohort details}
In order to train and validate our proposed network, we have used three different publicly available open-source cohorts of CXRs images, namely, the \textbf{Pediatric CXRs dataset} to detect pneumonia vs healthy cases \cite{KERMANY2018} (source-1), the \textbf{IEEE COVID-19 CXRs dataset} \cite{cohen2020covidProspective} (source-2), and the \textbf{Tuberculosis CXRs from Shenzhen Hospital x-ray dataset} {\cite{tr}} (source-3). It is important to mention that there were no multi-label cases, such as pneumonia and COVID-19 findings in the same patient, in any of these datasets.

In order to demonstrate the adaptability of our model in multi-class datasets, we created four different datasets, namely DXR1, DXR2, DXR3, and DXR4. The DXR1 dataset was based on source-1 cohort with 3883 pneumonia images and 1350 healthy images. This dataset is a binary classification dataset to detect pneumonia and healthy cases. The source-1 cohort is collected based on paediatric populations. Next, in DXR2, we have trained and tested the models for classification of COVID-19, pneumonia, and healthy patients in the IEEE COVID-19 x-rays dataset (source-2) with 69 COVID-19 images, 79 pneumonia images, and 79 healthy cases. In the third dataset (DXR3) of our study, we have trained and validated our network on source-2 and the tuberculosis (TB) cases of Shenzhen Hospital x-ray dataset (source-3) to detect TB, COVID-19, pneumonia, and healthy cases. As the source-3 had more than 300 CXR images for both TB and healthy classes, the combination of the two sources would end up with an unbalanced dataset. Thus, we randomly selected 79 tuberculosis images from source-3 and 69 COVID-19 images, 79 pneumonia images, and 79 healthy cases from source-2, in order to generate the DXR3. In the DXR4 dataset of our study, we have trained and validated our network on a combination of source-1, source-2, and source-3 to detect TB, COVID-19, pneumonia, and healthy cases. Only in this case, we mixed the pediatric and adult patients populations of the three sources, in order to test the robustness of our proposed model in multi-class dataset with a variation of the patient's age. To avoid the bias effects, we created the dataset with randomly selected balanced number of images. In the healthy class, we included 110 images from each source to generate a total of 330 healthy cases. In the pneumonia class, we included 79 images from source-2 and 221 images from source-1 to generate 300 pneumonia images. Finally, 310 tuberculosis images source-3 and 69 COVID-19 images from source-2 were included to prepare the final DXR4 dataset. To the best of our knowledge, there was no other COVID-19 CXR open-source dataset available, in order to balance the number of images in the COVID-19 class. Figure~\ref{cxr} depicts a sample of the CXR scans from the healthy, COVID-19, pneumonia, and tuberculosis patients, as determined by expert radiologists. Summarizing the four different cases:
\begin{itemize}
\item DXR1: $3883$ pneumonia and $1350$ healthy cases \cite{KERMANY2018}.
\item DXR2: $69$ COVID-19 images, $79$ pneumonia images, and $79$ healthy cases \cite{cohen2020covidProspective}.
\item DXR3: $69$ COVID-19 images, $79$ pneumonia images, $79$ tuberculosis images, and $79$ healthy cases \cite{cohen2020covidProspective,tr}.
\item DXR4: $69$ COVID-19 images, $300$ pneumonia images, $310$ tuberculosis images, and $330$ healthy cases \cite{KERMANY2018,cohen2020covidProspective,tr}.
\end{itemize}
Here, the DXR4 is simply an extended version of DXR3, generated using more images from the datasets of \cite{KERMANY2018, cohen2020covidProspective,tr}.

\subsection{Cohort's pre-processing image analysis}
Image analysis techniques have been applied on all slices to reduce the effect of noise and increase the signal-to-noise ratio (SNR). We have used noise filters such as binomial deconvolution, Landweber deconvolution \cite{Vonesch2008}, and curvature anisotropic diffusion image filters \cite{Perona1990} to reduce noise in the images. We have normalized the images by subtracting the mean value from each image and dividing by its standard deviation. Finally, we have used data augmentation techniques including rotation (rotation around the center of image by a random angle in the range of $-15^{\circ} $ to $ 15^{\circ}$), width shift range (width shift of image by up to 20~pixels), height shift range (height shift of image by up to 20~pixels), and ZCA whitening (add noise in each image) \cite{ko}.

\subsection{Hyper-parameters initialization}
After random shuffling, each dataset has been partitioned into 70\% and 30\% of the total CXR images using the repeated random subsampling validation technique (also known as the Monte Carlo cross-validation split), before training and testing the models, respectively. We have used the categorical cross-entropy as cost function. The loss function is optimized using the stochastic gradient descent (SGD) method with learning rate of $0.001$ and with $30$ epochs (the models converged after 20-25 epochs). We have applied transfer learning techniques on the ResNet-50 and DenseNet-121 networks using the ImageNet dataset \cite{imagenet} (\url{http://www.image-net.org}). It consists of over $14$ million images and the task is to classify the images into one of almost $22,000$ different categories (cat, sailboat, etc.).

\subsection{Software}
The code developed in this study is written in the Python programming language using Keras/TensorFlow (Python) libraries. For training and testing of the deep learning networks, we have used an NVIDIA cluster, with $4$ GPUs and $64$ GB RAM memory. The code implementation is available on a public repository with url: \url{https://github.com/team-globs/COVID-19_CXR}.

\section{Performance analysis and discussions}
\label{sec:results}
This section presents the performance of our proposed DenResCov-19 network, along with a quantitative performance comparison with established DL networks, on four different datasets. The underlying reason behind choosing the DenseNet-121 and ResNet-50 networks in our study is that we wanted to combine the advantages of both networks to develop a new network with well-balanced AUC-ROC and F1 metric values. VGG-16 is a network with relatively faster training time and, in the majority of cases, it has very good AUC-ROC, but comparatively poor F1-value. Hence, we wished to check if the performance of our network is superior enough from VGG, to compensate for the relatively slower training procedure.  We preferred to choose ResNet-50 as a well-balanced choice regarding the training time and accuracy of the network, since ResNet is very fast in low layers (such as ResNet-18), but the accuracy improves as the layers of the structure increase (50, 110 etc.). The same approach was followed for the DenseNet too. 

In addition, in the study presented in \cite{nature}, the DL structures with superior performance in classification were determined as ResNet, DenseNet, AlexNet, Inception, VGG, and SqueezeNet. Among these networks, the most superior AUC-ROC value in COVID 19 image data collection and CXR cohort were the ResNet-50, DenseNet-161, VGG-19, and AlexNet. Regarding the Area Under the Precision Recall Curve and Sensitivity and Specificity, the best networks were the ResNet-50, DenseNet-161, VGG-16, and Alex-Net. Since ResNet-50 and DenseNet-161 presented satisfactory performance in the majority of the cases, we preferred to consider them as the benchmark networks. However, instead of DenseNet-161, we used the DenseNet-121 due to its significantly less computation time during training. 

\subsection{Evaluating the classification performance}
\begin{table*}
\caption{Comparative performance metrics of the different deep learning networks performing classification of  pneumonia, TB, COVID-19, and healthy cases. \textbf{*Boldface} indicates the best metric among the networks.}
\centering
\begin{tabular}{ |p{2.5cm}||p{2.5cm}|p{2.5cm}|p{2.5cm}||p{2.5cm}| }
    \toprule
 \multicolumn{5}{|c|}{DXR1 dataset: pneumonia and healthy \cite{KERMANY2018}} \\
\cmidrule(r){1-5}
 Metric     & DenResCov-19 & DenseNet-121 & ResNet-50 & Inception-V3 \\
 \midrule
Recall (\%) & \textbf{98.12}$^*$  & 97.80 & 97.71 & 93.32 \\
Precision (\%)  & \textbf{98.31}$^*$  & 94.62 & 95.01 & 90.10 \\
AUC-ROC (\%) & \textbf{99.60}$^*$ & 99.10 & 98.95	& 92.80 \\
F1 (\%)  &  \textbf{98.21}$^*$ & 96.27 & 96.34	& 91.68 \\
\bottomrule
\end{tabular}

\begin{tabular}{ |p{2.5cm}||p{2.5cm}|p{2.5cm}|p{2.5cm}||p{2.5cm}|}
\toprule
\multicolumn{5}{|c|}{DXR2 dataset: COVID-19, pneumonia and healthy \cite{cohen2020covidProspective}} \\
\midrule
 Metric     & DenResCov-19 & DenseNet-121 & ResNet-50 & VGG-16 \\
\midrule
Recall (\%) &  89.38 & 83.54 & 83.53 & \textbf{99.83}$^*$ \\
Precision (\%)  & \textbf{85.28}$^*$   & 77.45	& 73.35	 & 33.38 \\
AUC-ROC (\%) & \textbf{96.51}$^*$ & 93.2 & 92.39	&  50.07\\
F1 (\%)  & \textbf{87.29}$^*$   & 80.37 & 78.11	&  49.51 \\
\bottomrule
\end{tabular}

\begin{tabular}{ |p{2.5cm}||p{2.5cm}|p{2.5cm}|p{2.5cm}||p{2.5cm}|}
\toprule
\multicolumn{5}{|c|}{DXR3 dataset: COVID-19, pneumonia, tuberculosis and healthy \cite{cohen2020covidProspective,tr}} \\
\midrule
 Metric     & DenResCov-19 & DenseNet-121 & ResNet-50 & VGG-16 \\
\midrule
Recall (\%) &59.28 & 57.71	& 56.66 & \textbf{66.53}$^*$  \\
Precision (\%)  &  \textbf{79.56}$^*$ & 74.87  &  74.00	 & 26.53  \\
AUC-ROC (\%) & \textbf{91.77}$^*$ & 89.49 & 92.12	& 53.11 \\
F1 (\%)  & \textbf{68.09}$^*$ & 65.17 & 64.17 & 38.00  \\
\bottomrule
\end{tabular}
\begin{tabular}{ |p{2.5cm}||p{2.5cm}|p{2.5cm}|p{2.5cm}||p{2.5cm}|}
\toprule
\multicolumn{5}{|c|}{DXR4 dataset: COVID-19, pneumonia, tuberculosis and healthy \cite{KERMANY2018,cohen2020covidProspective,tr} } \\
\midrule
 Metric     & DenResCov-19 & DenseNet-121 & ResNet-50 & VGG-16 \\
\midrule
Recall (\%) &69.7 & 62.70	& 62.00 & \textbf{93.69}$^*$ \\
Precision (\%)  & \textbf{82.90}$^*$  & 79.35  & 78.60  	 & 27.17 \\
AUC-ROC (\%) & \textbf{95.00}$^*$ & 91.00 & 93.21	& 54.99 \\
F1 (\%)  &  \textbf{75.75}$^*$  & 70.07 & 69.51	& 42.13  \\
\bottomrule
\end{tabular}
\label{tab3}
\end{table*}

As explained in Section 4.1, we have created four different CXR image collections to evaluate the performance of the models in binary and multiclass classification. Table~\ref{tab3} summarizes the metrics for the different networks and datasets. Our initial hypothesis that our network DenResCov-19 will have more balanced AUC-ROC and F1 measurements compared to  the DenseNet-121 and ResNet-50 networks, has been verified in all four datasets.

In particular, DenResCov-19 has AUC-ROC of $99.60$, $96.51$, and $95.00\%$, contrary to the $98.95$, $92.12$, and $93.21\%$ of ResNet-50 and $99.10$, $93.20$, and $91.00\%$ of DenseNet-121 for the DXR1, DXR2, and DXR4 datasets, respectively. In addition, DenResCov-19 has F1 values of $98.21$, $87.29$, and $75.75\%$, contrary to the $96.34$, $78.11$, and $69.51\%$ of ResNet-50 and $96.27$, $80.37$, and $70.07\%$ of DenseNet-121 for the DXR1, DXR2, and DXR4 datasets, respectively. Our network has achieved more than $98\%$ in all metrics in the binary label classification (pneumonia or healthy) of the DXR1 dataset. With the exception of the recall values in DXR2 and DXR4 datasets of VGG-16, our approach outperforms all other networks for all four metrics in all four datasets.

From the results presented in Table~\ref{tab3}, it is clear that as the number of label classes increases, the accuracies of evaluation metrics decrease. In our DenResCov-19 network, the recall value of $98.12\%$ in DXR1 has decreased in DXR2, DXR3, and DXR4 datasets with a variation between $59.28\%$ to $89.38\%$. In a similar way, the precision value has reduced from $98.31\%$ to $79.56-85.28\%$, AUC-ROC from $99.60\%$ to $91.77-96.51\%$, and the F1-value from $98.21\%$ to $68.09-87.29\%$. However, as previously discussed, the results or our network are still better than the state-of-the-art networks. It should also be noted that the metric results in DXR4 dataset are better than the results in DXR3, although the numbers of label classes in two datsets are the same (COVID-19, pneumonia, tuberculosis, and healthy). This happens as the number of training data has increased from almost $80$ images to almost $300$ images per class (except for the COVID-19 cases, which remains at 69). It is worth mentioning that, since the number of labelled COVID-19 X-ray images is very limited ($69$ images), it has affected the quantitative results of both precision and recall values in DXR2, DXR3, and DXR4 datasets. Incorporation of additional labelled data in future would significantly improve the performance with respect to these two indices.

\subsection{Evaluating the cross validation results }

For any classification task, it is very important to minimize the bias effects generated from a fixed validation scheme (70\% training, 30\% testing). Thus, we have compared the three networkd (DenseNet-121, ResNet-50, and DenResCov-19) in DXR4 dataset for classification over four randomly shuffled fixed ratio validation schemes (also  known as Monte Carlo cross-validation method). For each cross-validation set, we have calculated the F1 and AUC-ROC metrics and the `micro', `macro', and `weighted' versions of the indices. The `micro' version is calculated by counting the total number of true positives, false negatives, and false positives. The `macro' version computes the metric for each class and finds their unweighted means. The `weighted' version measures the metric for each class and determines their weighted means.

Table \ref{tab33} summarises the results of four different cross-validations in the DXR4 dataset for the DenseNet-121, ResNet-50, and DenResCov-19 networks. DenResCov-19 achieves the highest score in all average, higher, and lower values of the metrics. DenseNet-121 has higher recall and precision average values (62.7, 79.3\% against 62.0, 78.6\%) and lower AUC-ROC (91.0 against 93.2\%) as compared the ResNet-50 network. 

\begin{table*}
\caption{Quantitative evaluation metrics for four cross-validation cases on the DXR4 dataset, and the resulting average and standard deviation. Superscript max/min indicates the highest/lowest score among the cross validation sets.}
\centering
\begin{tabular}{ |p{3cm}||p{2.25cm}|p{2.25cm}|p{2.25cm}||p{2.25cm}||p{2.0cm}| }
\toprule
\multicolumn{6}{|c|}{Classification performance of ResNet-50 in DXR4 dataset: COVID-19, pneumonia, tuberculosis and healthy \cite{KERMANY2018,cohen2020covidProspective,tr}  } \\
\midrule
 Metric (\%)  & Cross validation \#1 & Cross validation \#2 & Cross validation \#3 &  Cross validation \#4 &Average value\\
\midrule

Recall  & {62.7}$^{\max}$ & 61.9 & 62.4 & {61.5}$^{\min}$ & \textbf{62.0 $\pm$ 0.5} \\
Precision    & {81.0}$^{\max}$ & 78.9 & {77.1}$^{\min}$ & 77.6 & \textbf{78.6 $\pm$ 1.7} \\
AUC-ROC   & {94.1}$^{\max}$  & 93.2  & 92.7 & {92.6}$^{\min}$ & \textbf{93.2 $\pm$ 0.7}\\
AUC-ROC macro  & 89.9  & {89.0}$^{\min}$ & {91.0}$^{\max}$ & 89.5 & \textbf{89.9 $\pm$ 0.9}\\
AUC-ROC micro  &  {88.1}$^{\min}$ & 88.8 & {90.4}$^{\max}$ & 88.8 & \textbf{89.0 $\pm$ 1.0}\\
AUC-ROC weighted  & {87.3}$^{\min}$  & 88.4 & {89.6}$^{\max}$ & 88.2 & \textbf{88.4 $\pm$ 0.9} \\
F1  & {70.7}$^{\max}$ & 69.9 & 69.0 & {68.6}$^{\min}$ & \textbf{69.5 $\pm$ 0.9}\\
F1 macro  & {70.7}$^{\max}$ &  70.0 & 69.1 & {68.6}$^{\min}$ & \textbf{69.6 $\pm$ 0.9} \\
F1 micro   & {70.5}$^{\max}$  & 69.8 & 68.8 & {68.4}$^{\min}$ & \textbf{69.4 $\pm$ 0.9}\\
F1 weighted   & {70.7}$^{\max}$  & 70.1 & 69.0 & {68.7}$^{\min}$ & \textbf{69.6 $\pm$ 0.9}\\
\bottomrule

\end{tabular}
\begin{tabular}{ |p{3cm}||p{2.25cm}|p{2.25cm}|p{2.25cm}||p{2.25cm}||p{2.0cm}|}
\toprule
\multicolumn{6}{|c|}{Classification performance of DenseNet-121 in DXR4 dataset: COVID-19, pneumonia, tuberculosis and healthy \cite{KERMANY2018,cohen2020covidProspective,tr} } \\
\midrule

 Metric  (\%)  & Cross validation \#1 & Cross validation \#2 & Cross validation \#3 & Cross validation \#4  & Average value\\
\midrule

Recall  & {63.3}$^{\max}$ & 62.4 & {62.3}$^{\min}$ & 62.9 & \textbf{62.7 $\pm$ 0.5} \\
Precision    & 80.1 & {80.8}$^{\max}$ & 79.3 & {76.8}$^{\min}$ & \textbf{79.3 $\pm$ 1.7} \\
AUC-ROC  & {93.8}$^{\max}$ & 91.2 & {89.0}$^{\min}$ & 89.8 & \textbf{91.0 $\pm$ 2.1} \\
AUC-ROC macro  & {90.1}$^{\min}$ & {92.5}$^{\max}$ & 90.2  & 91.5 & \textbf{91.1 $\pm$ 1.1} \\
AUC-ROC micro   & 89.2 & {91.2}$^{\max}$ &  {88.8}$^{\min}$ & 89.2 & \textbf{89.6 $\pm$ 1.1} \\
AUC-ROC weighted  & 88.7  & {90.6}$^{\max}$ &  87.6 & {87.2}$^{\min}$ & \textbf{88.5 $\pm$ 1.5}\\
F1    & {70.7}$^{\max}$ & 70.4 &  69.8 & {69.2}$^{\min}$ & \textbf{70.0 $\pm$ 0.6}\\
F1 macro   & {70.0}$^{\max}$ & 69.6 & 69.0 & {68.3}$^{\min}$ & \textbf{69.3 $\pm$ 0.7}\\
F1 micro   & {70.7}$^{\max}$ & 70.4 &  69.8 & {69.2}$^{\min}$ & \textbf{70.0 $\pm$ 0.6}\\
F1 weighted   & {70.4}$^{\max}$ & 69.9 & 69.6 & {69.0}$^{\min}$ & \textbf{69.8 $\pm$ 0.6}\\

\bottomrule
\end{tabular}
\begin{tabular}{ |p{3cm}||p{2.25cm}|p{2.25cm}|p{2.25cm}||p{2.25cm}||p{2.0cm}|}
\toprule
\multicolumn{6}{|c|}{Classification performance of DenResCov-19 in DXR4 dataset: COVID-19, pneumonia, tuberculosis and healthy \cite{KERMANY2018,cohen2020covidProspective,tr} } \\
\midrule
 Metric  (\%)  & Cross validation \#1 & Cross validation \#2 & Cross validation \#3 & Cross validation \#4 & Average value\\
\midrule
Recall  & 70.0 & {71.0}$^{\max}$ & {67.0}$^{\min}$ & 70.7  & \textbf{69.7 $\pm$ 1.8}\\
Precision    & {80.0}$^{\min}$ & 83.0 & {86.0}$^{\max}$ & 82.6 & \textbf{82.9 $\pm$ 2.4}\\
AUC-ROC   & {93.9}$^{\min}$  & 95.0 & {96.0}$^{\max}$ & 95.2 & \textbf{95.0 $\pm$ 0.8}\\
AUC-ROC macro   & {94.7}$^{\min}$  & 94.7 & {94.7}$^{\max}$ & 94.8 & \textbf{95.6 $\pm$ 0.1}\\
AUC-ROC micro   & {93.9}$^{\min}$  & 94.4 & {98.2}$^{\max}$ & 93.9 & \textbf{95.1 $\pm$ 2.1}\\
AUC-ROC weighted   & {93.3}$^{\min}$ & 94.1 & {98.0}$^{\max}$ & 93.6  & \textbf{94.7 $\pm$ 1.8}\\
F1   & {75.0}$^{\min}$ &  {76.5}$^{\max}$ & 75.3 & 76.2 & \textbf{75.8 $\pm$ 0.7} \\
F1 macro  & 76.2 & {77.6}$^{\max}$ & {76.1}$^{\min}$ & 77.1 & \textbf{76.7 $\pm$ 0.7}\\
F1 micro  & {74.9}$^{\min}$ & {76.3}$^{\max}$ & 75.3 & 76.2 & \textbf{75.6 $\pm$ 0.7}\\
F1 weighted  & {75.0}$^{\min}$  & {76.5}$^{\max}$ & 75.3 & 76.2 & \textbf{75.7 $\pm$ 0.7}\\
\bottomrule
\end{tabular}
  \label{tab33}
\end{table*}

Figure~\ref{cx} presents the ROC curves for multi-class classification by ResNet-50, DenseNet-121, and DenResCov-19 networks. The ROC curves are computed in four different cross-validation cases in DXR4 dataset. Based on these figures, it is clear that the true/false positive rate and the ROC curves' results of the DenResCov-19 network (Fig.~\ref{cx} third row) are much better in all classes, as compared to the other two networks (Fig.~\ref{cx} first and second rows). From the results presented in the Fig.~\ref{cx}, we can find the average AUC-ROC values of the four classes for DenseNet-121, ResNet-50, and DenResCov-19 networks. The average AUC-ROC values of TB, COVID-19, healthy, and pneumonia classes for ResNet-50 are 84.8, 82.5, 92.3, 87.1\%, while the same for DenseNet-121 are 87.3, 83.1, 90.8, 89.7\% and for DenResCov-19 are 94.7, 92.6, 96.4, 95.3\%, respectively. Hence, the DenseNet-121 achieves improved true/false positive rate and AUC-ROC values as compared to the ResNet-50 (except for the healthy class). On the other hand, the performance of DenResCov-19 is higher in all average AUC-ROC values of TB, COVID-19, healthy, and pneumonia classes compared to both ResNet-50 and DenseNet-121.

\begin{figure*}
 \centering\includegraphics[width=1.25\textwidth]{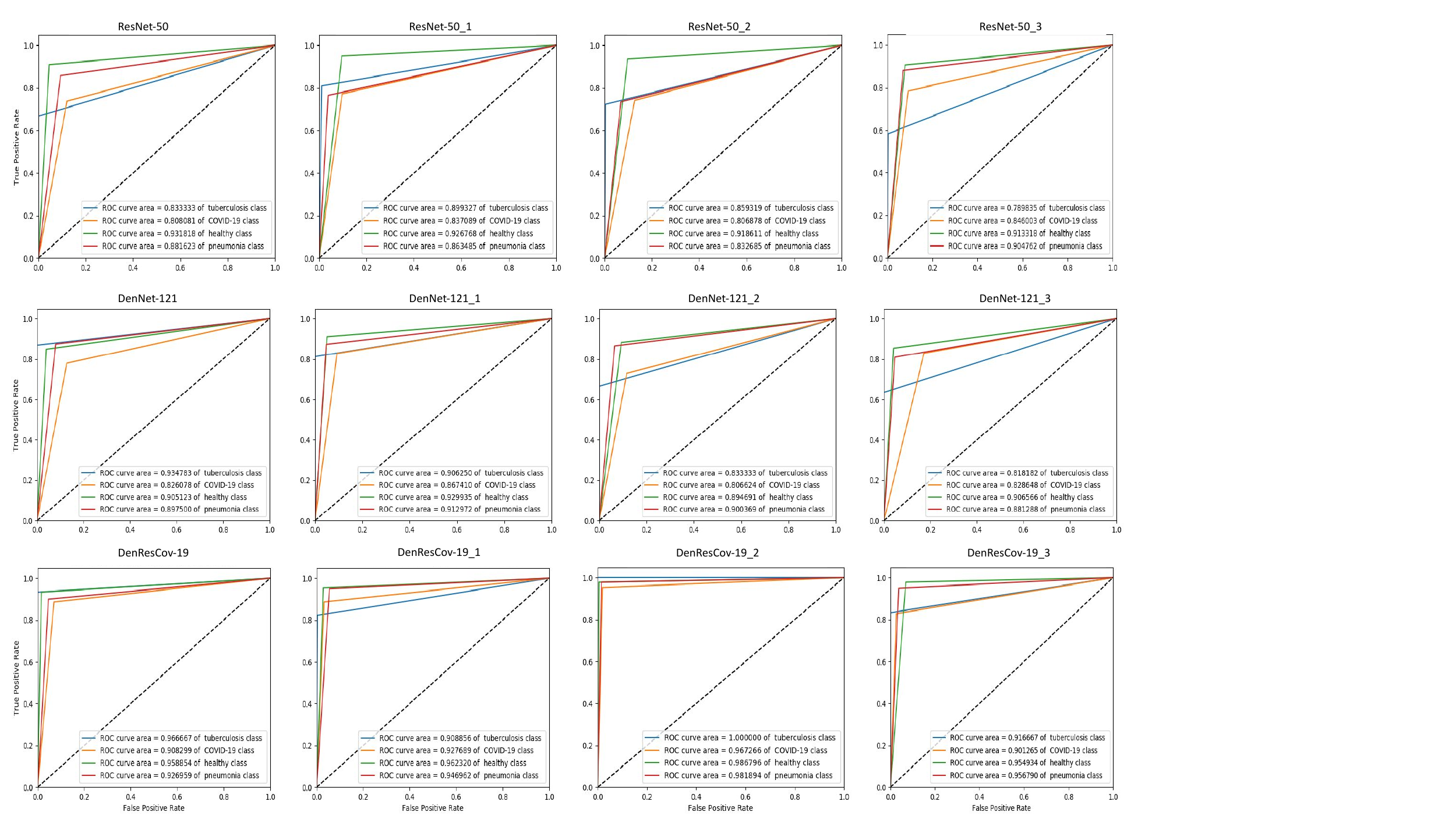}
  \caption{The ROC curves for the four cross-validation cases over DXR4 dataset. Top to bottom: ResNet-50, DenseNet-121, and DenResCov-19.}
  \label{cx}
\end{figure*}  
\begin{figure*}
\centerline{
\includegraphics[width=.333\textwidth]{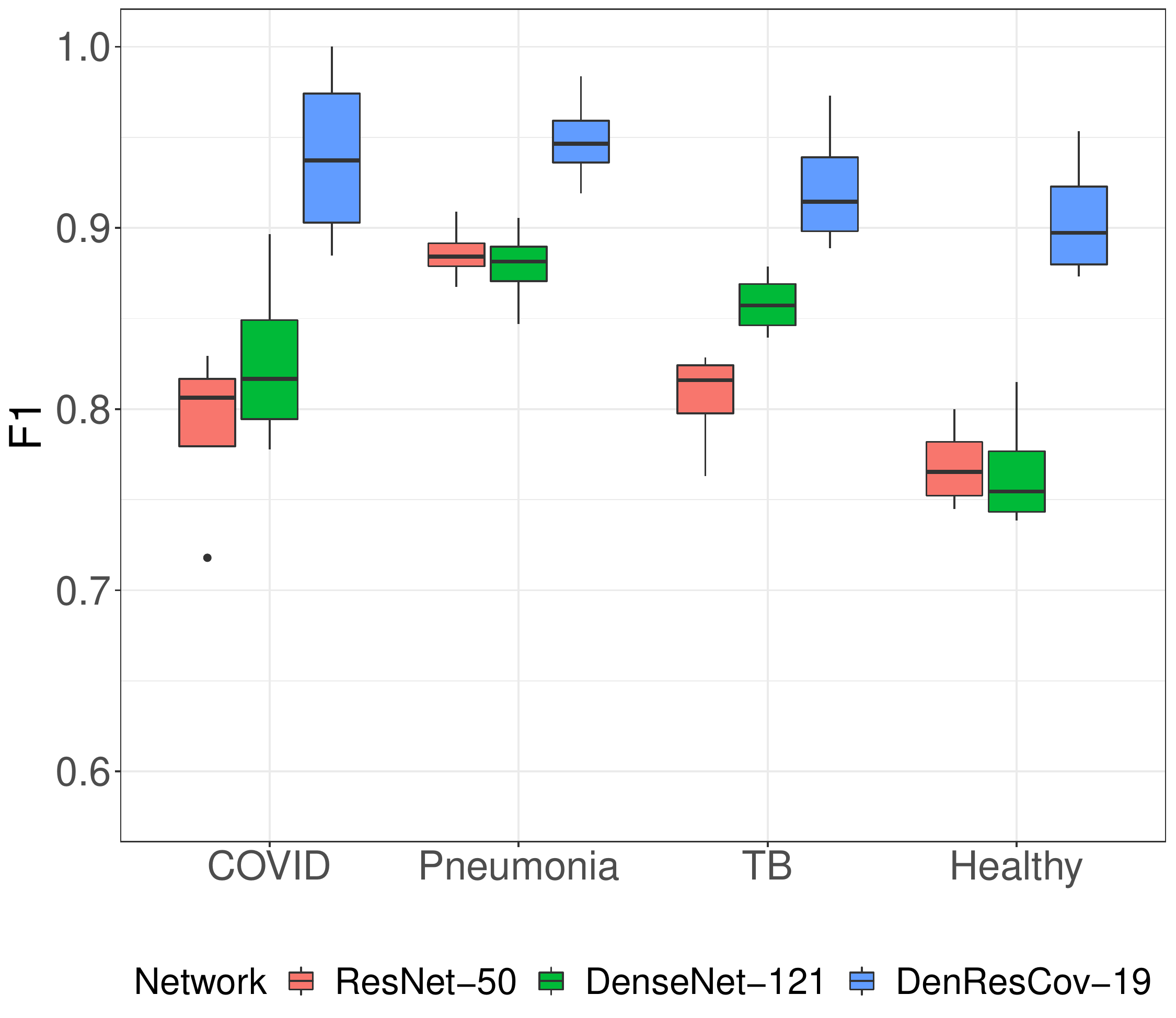}
\includegraphics[width=.333\textwidth]{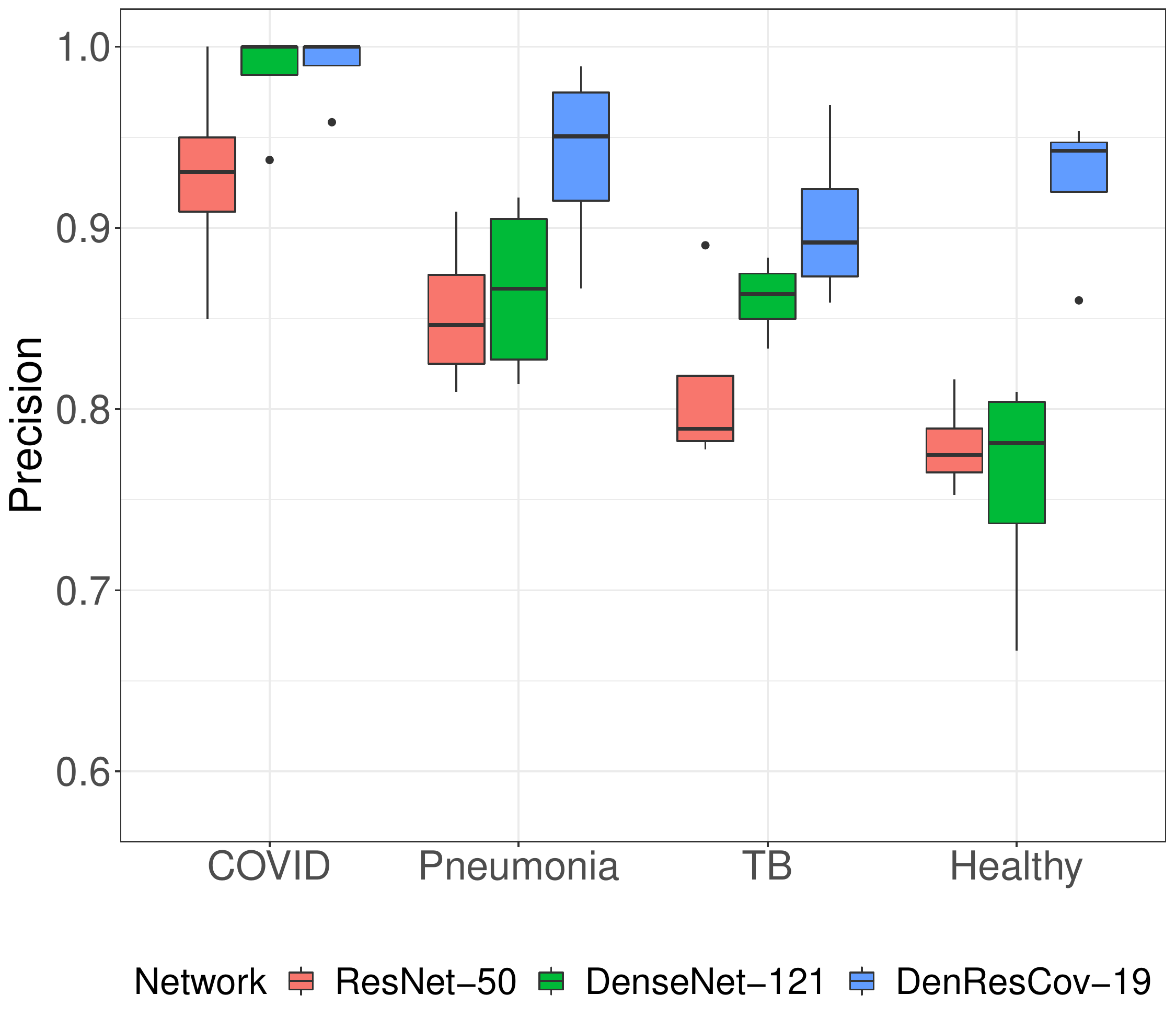}
\includegraphics[width=.333\textwidth]{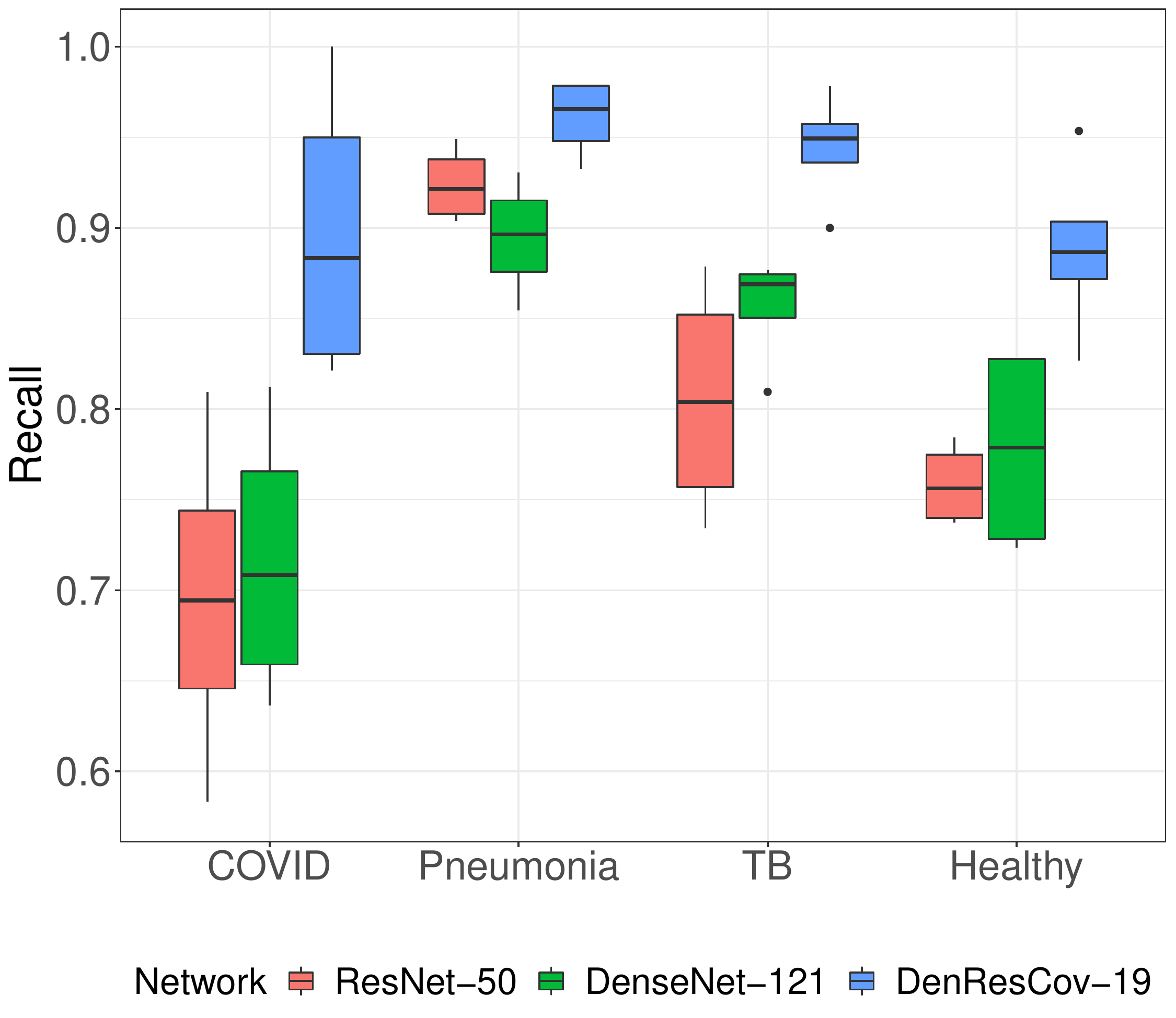}}
\caption{Boxplots for the quantitative performance analysis of three deep learning networks on DXR4 dataset for the classification of COVID-19, pneumonia, tuberculosis, and healthy patients.}
\label{boxplots_DXR4}
\vspace*{-0.15\baselineskip}
\end{figure*}

Figure~\ref{conf_matrix} presents the confusion matrices of multi-class classification by the ResNet-50, DenseNet-121, and DenRes- Cov-19 networks on DXR4 dataset (combined over four cross-validation cases). In ResNet-50 network, the COVID class has 69.2\% true positive and 30.8\% false negative predictions among the total number of positive cases, combined over four cross-validation iterations; while in the Pneumonia class, the network has 92.3\% true positive and 7.7\% false negative predictions. In the TB class, the network has 80.2\% true positive and 19.8\% false negative predictions, and in the Healthy class 75.8\% true positive and 24.2\% false negative predictions. On the other hand, the DenseNet-121 has in the COVID class 70.9\% true positive and 29.1\% false negative predictions, in the Pneumonia class 89.4\% true positive and 10.6\% false negative predictions, in the TB class 85.6\% true positive and 14.4\% false negative predictions, and in the Healthy class 77.5\% true positive and 22.5\% false negative predictions. In comparison, the DenResCov-19 network has in the COVID class 89.5\% true positive and 10.5\% false negative predictions, in the Pneumonia class 96.0\% true positive and 4.0\% false negative predictions, in the TB class 94.5\% true positive and 5.5\% false negative predictions, and in the Healthy class 88.5\% true positive and 11.5\% false negative predictions. Based on these evidences, we can infer that our proposed network results in higher true positive and lower negative false positive values as compared to the two established networks. Detailed results for the confusion matrices of individual cross validation cases of the three networks are provided in the supplementary material. 

The quantitative performance analysis of the Monte Carlo cross-validation experiment for ResNet-50, DenseNet-121, and DenResCov-19 networks over DXR4 dataset has been presented as box-plots in Fig.~\ref{boxplots_DXR4}. From the box-plots presented in Fig.~\ref{boxplots_DXR4}, it is clearly visible that the proposed DenRes- Cov-19 network achieves higher classification performance for all 4 classes, irrespective of the quantitative evaluation indices. For the statistical significance analysis of the classification performance of three networks in terms of the F1-score, precision, and recall values, the linear mixed model analysis has been adopted, where the four different classes, namely COVID-19, pneumonia, tuberculosis, and healthy patients, have been included as random effects in the linear mixed model. Applying the Kenward and Roger's method for the degrees of freedom of the \textit{t}-statistic \cite{KR1997} and the Tukey's method for pairwise comparisons \cite{Tukey1949}, we found that the proposed DenResCov-19 network achieves statistically significant classification performance over both DenseNet-121 and ResNet-50 networks in DXR4 dataset in terms of all three quantitative evaluation indices. In terms of F1-score, the DenResCov-19 attains significant p-values of $6.5E-09$ and $5.6E-11$ as compared to the DenseNet-121 and ResNet-50, respectively, while for the precision index, the p-values are measured as $0.0006$ and $2.3E-06$ as compared to the same two networks. Similarly with respect to the recall values, the proposed DenResCov-19 has attained significant p-values of $3.5E-06$ and $3.1E-07$ as compared to the DenseNet-121 and ResNet-50 networks, respectively.

\begin{figure*}
\centering
\begin{subfigure}[b]{0.325\textwidth}
\centering \includegraphics[width=\textwidth]{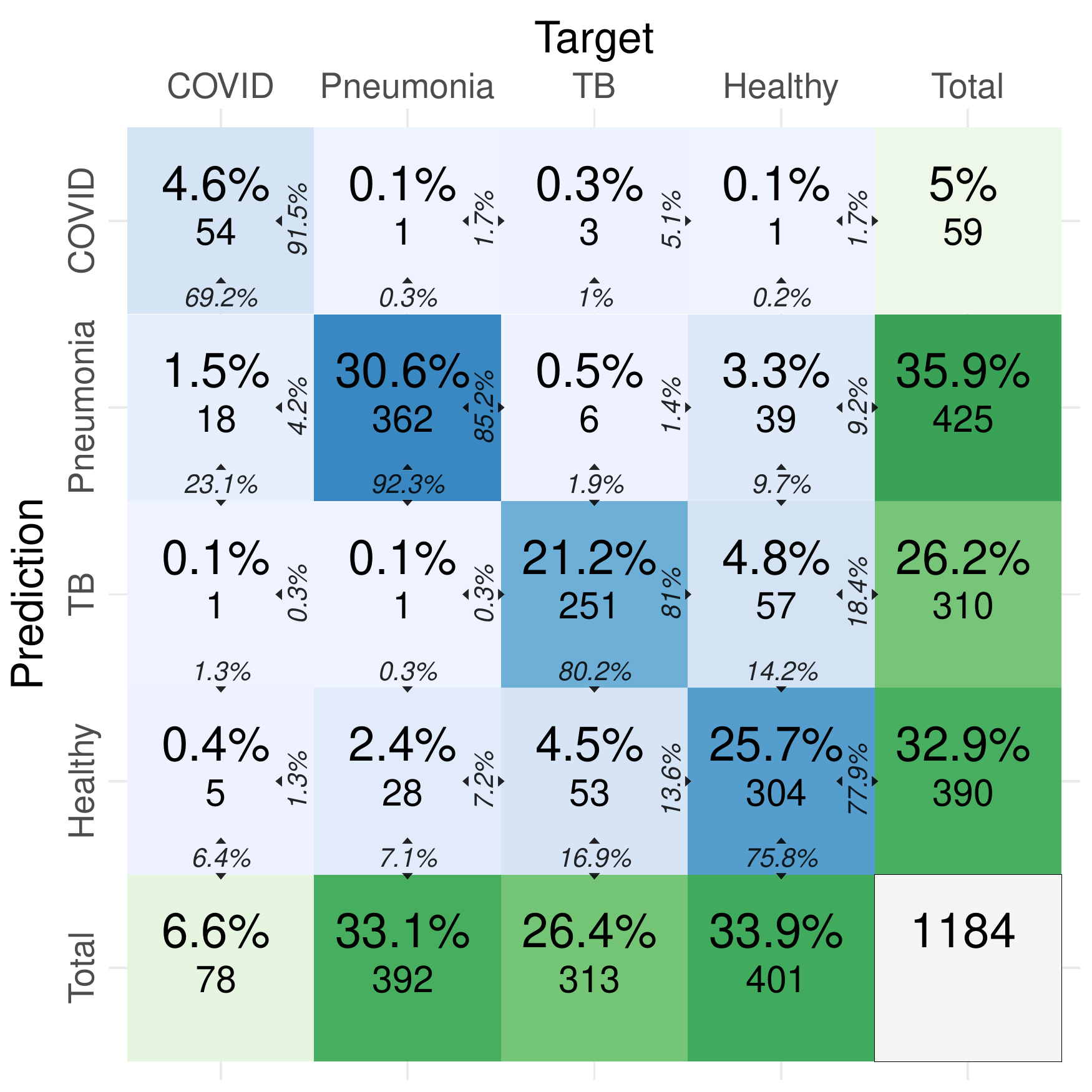}
\caption{ResNet-50}
\end{subfigure}
\begin{subfigure}[b]{0.325\textwidth}
\centering \includegraphics[width=\textwidth]{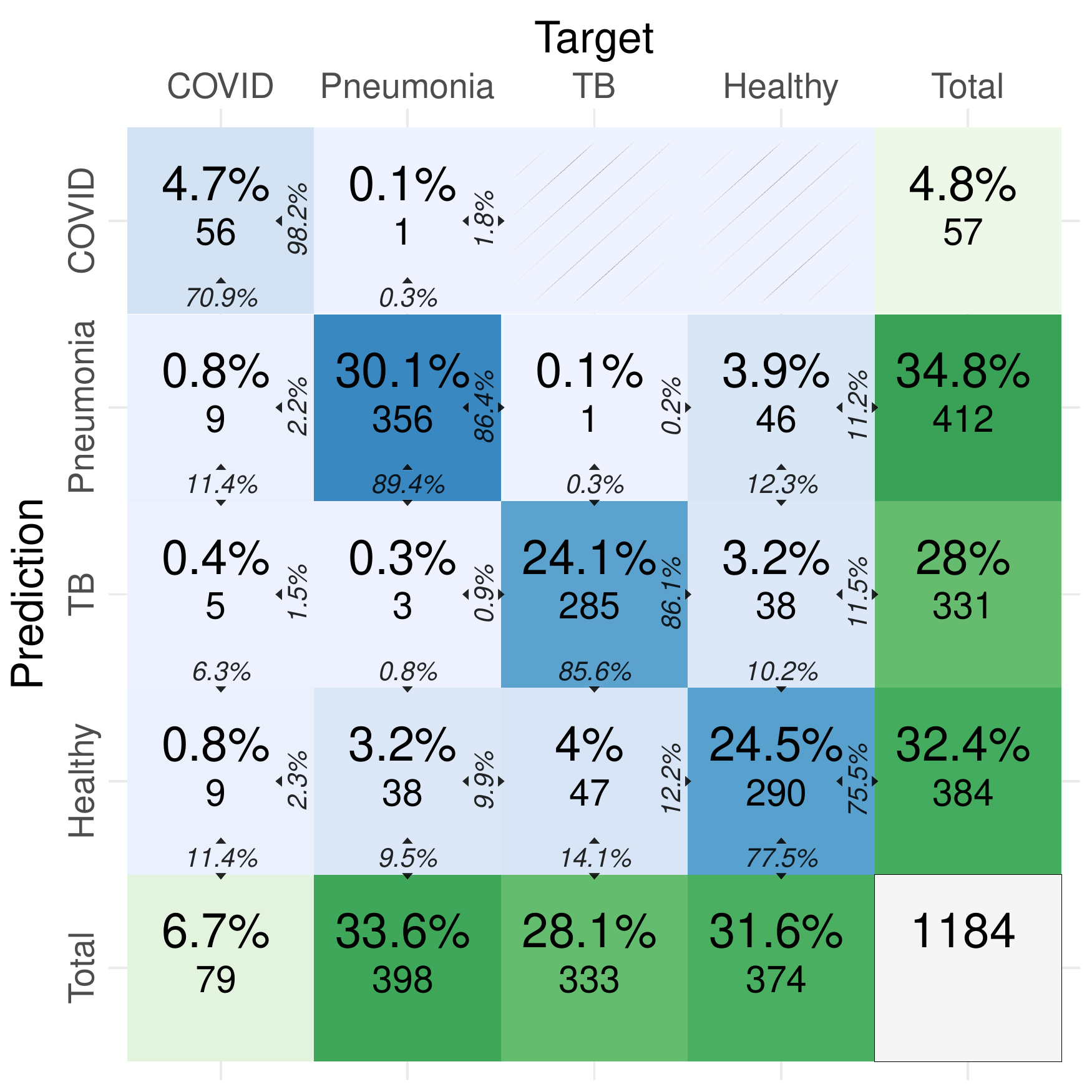}
\caption{DenseNet-121}
\end{subfigure}
\begin{subfigure}[b]{0.325\textwidth}
\centering \includegraphics[width=\textwidth]{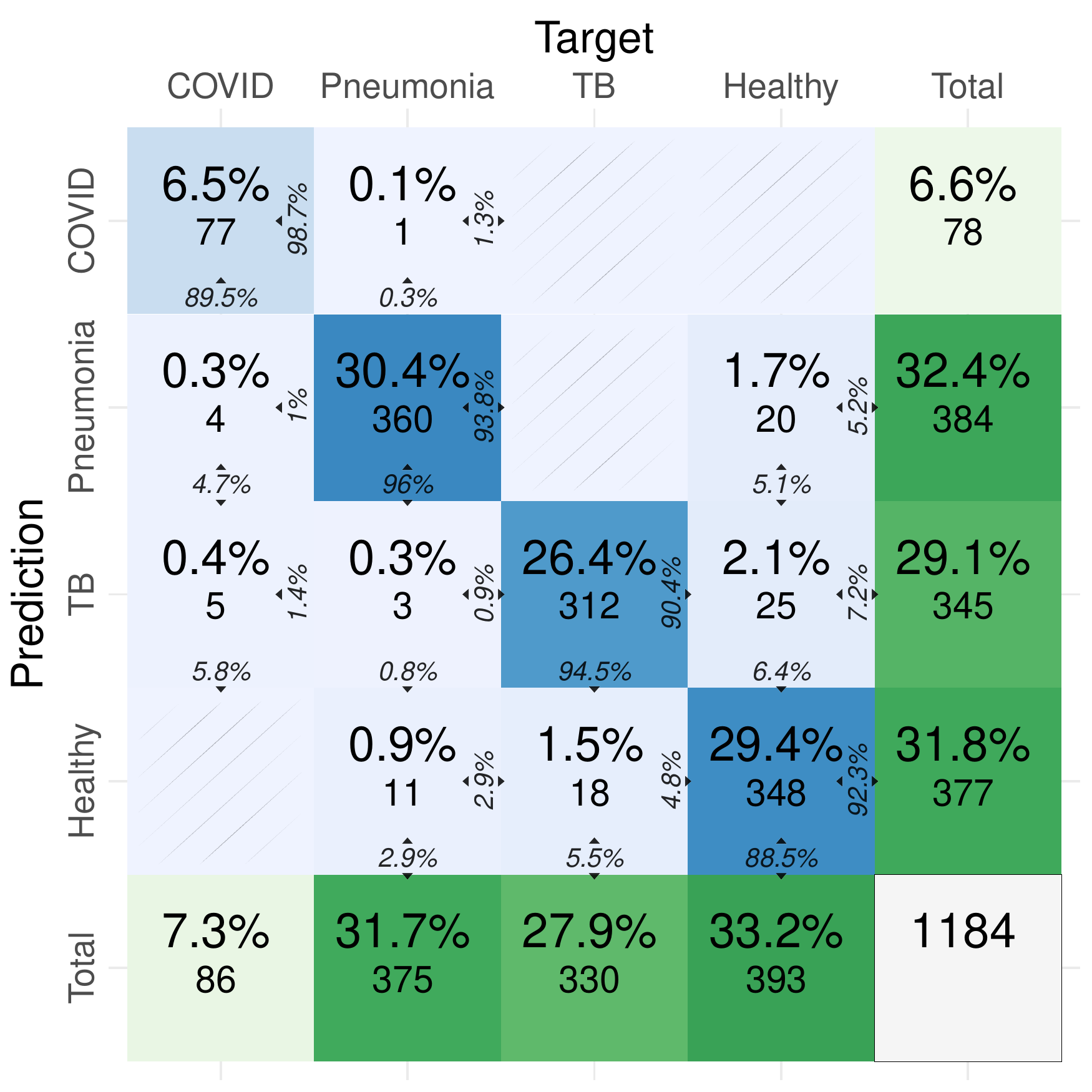}
\caption{DenResCov-19}
\end{subfigure}
\caption{Confusion matrices of the three deep learning networks on DXR4 dataset (combined over four cross-validation cases). Each blue-colored cell $(i,j)$ in the matrix denotes the number (and percentage) of cases in target class $i$ that has been classified as class $j$ during prediction. At the right edge of each cell, the percentage of cases in the cell with respect to prediction class $j$ is shown, while the bottom edge presents the percentage with respect to target class $i$. The last row and last column denote the total number (and percentage) of cases in the target and prediction classes, respectively.}
\label{conf_matrix}
\end{figure*}

\begin{figure*}
 \centering \includegraphics[width=.85\textwidth]{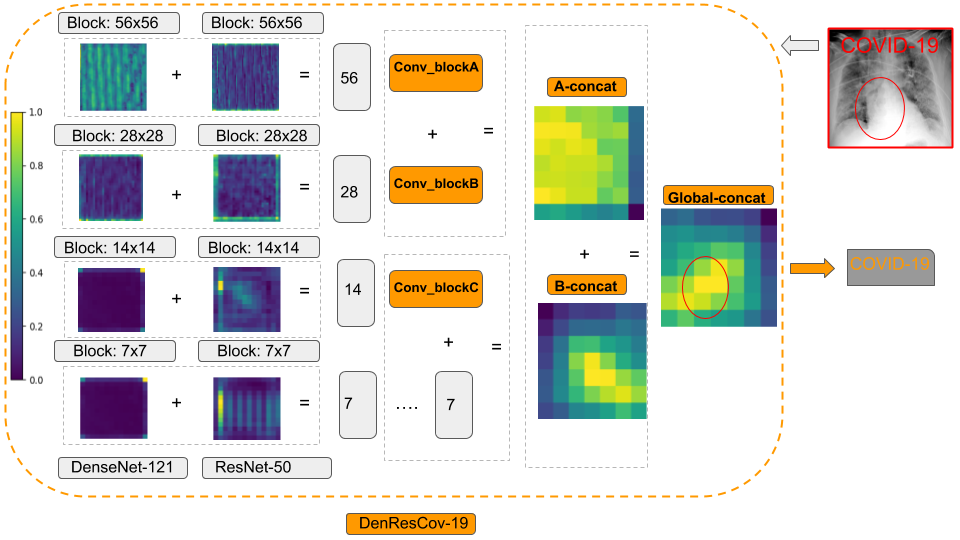}
 \caption{Main steps of the DenResCov-19 network to determine the classification decision, represented by heatmaps: The CXR image is the input of network, and the four blocks of ResNet-50 and DenseNet-121 outputs are then concatenated, creating four new heatmaps $56$, $28$, $14$, and $7$ (gray squares). The new heatmaps $56$, $28$, and $14$ initialize a convolution and average max-pool block layer (Conv-blockA, Conv-blockB, and Conv-blockC), and the exported images decreased to $7 \times 7$ pixels dimensions. Lastly, the network combines the Conv-blockA with the Conv-blockB outputs (A-concat) and the Conv-blockC with the new-heatmap $7$ (B-concat) which are concatenated in a Global-concat heatmap}
 \label{cxrw}
\end{figure*}

\begin{figure*}
 \centerline{\includegraphics[width=\textwidth]{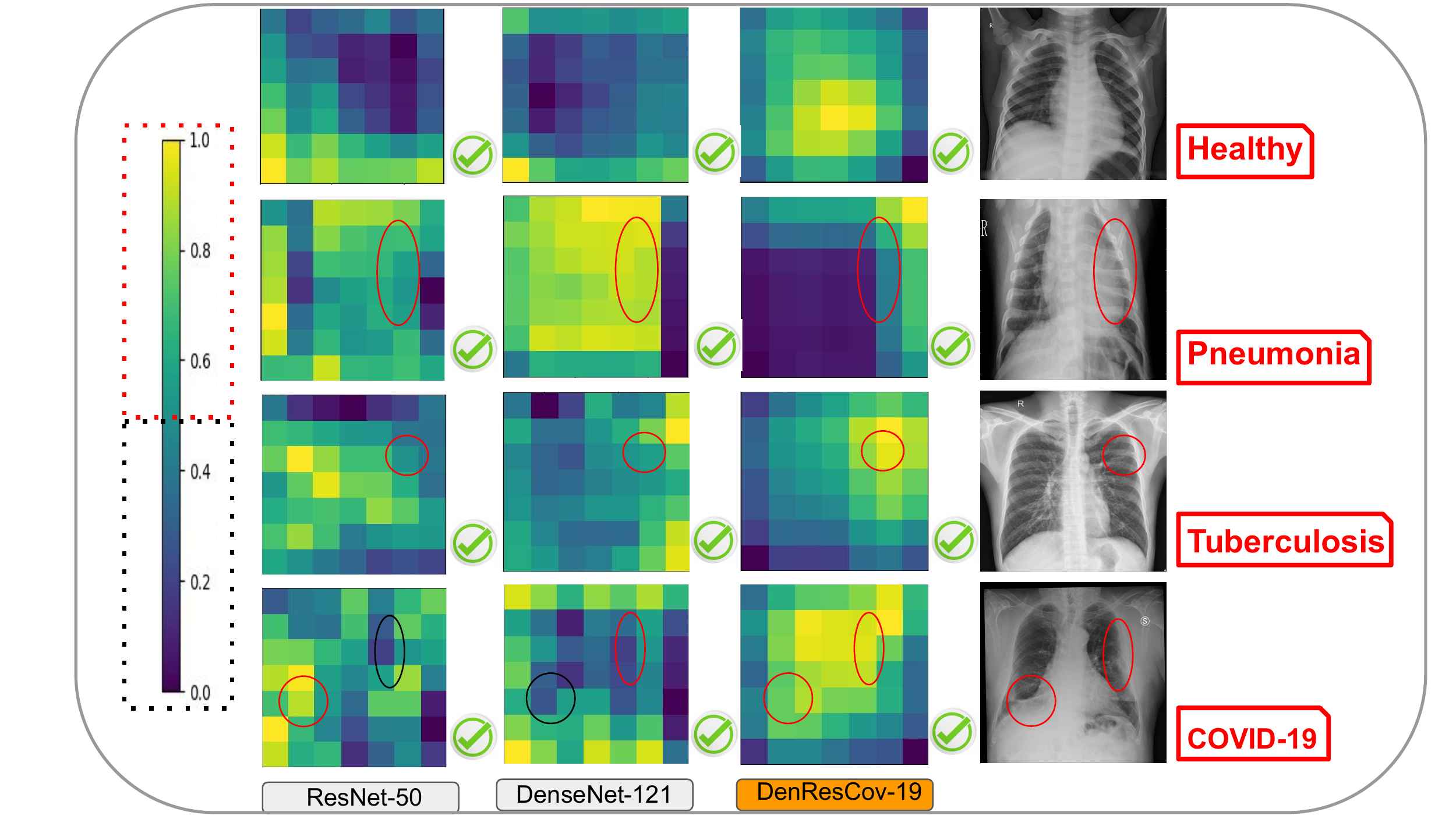}}
 \centerline{\includegraphics[width=\textwidth]{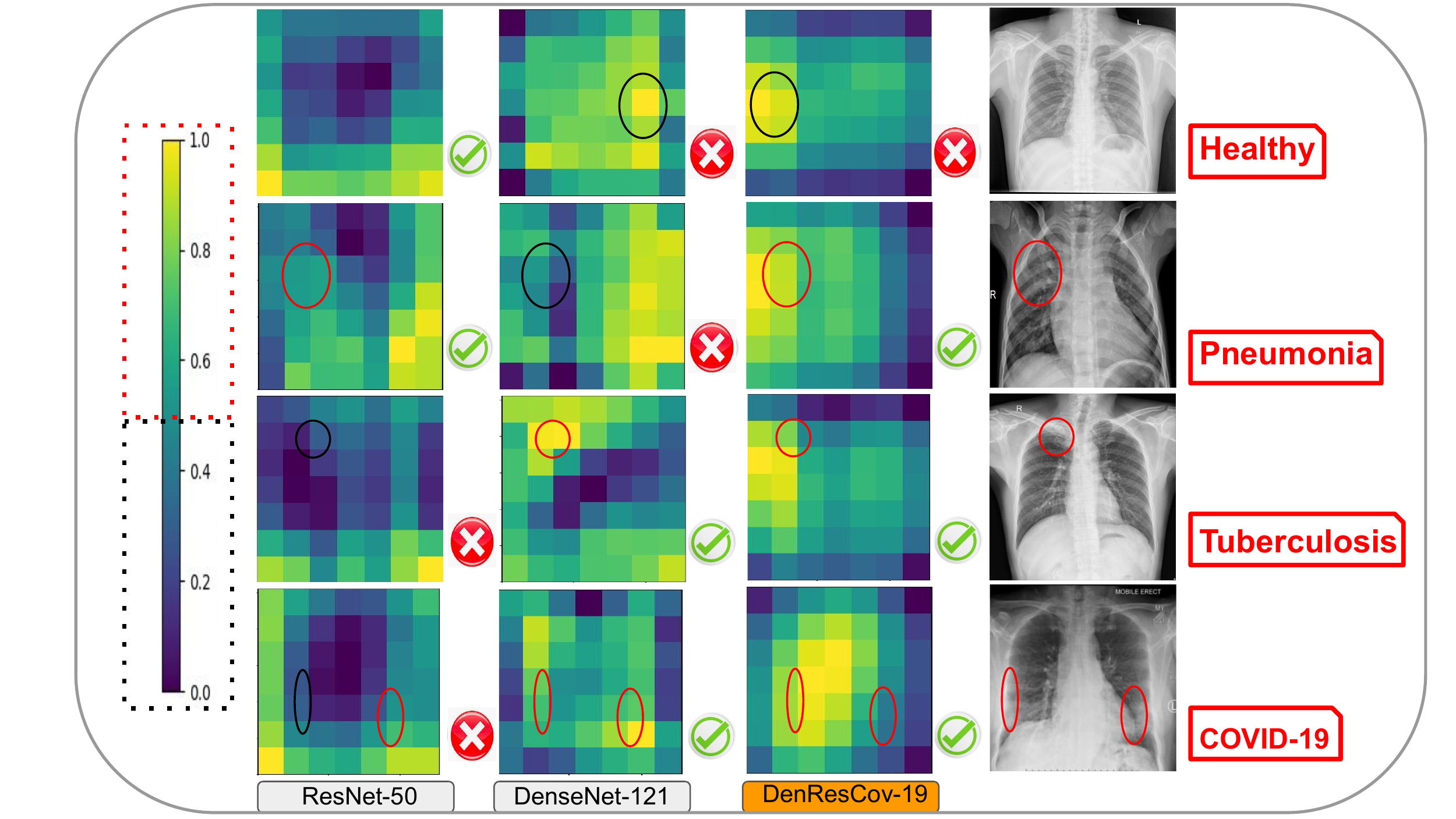}}
 \caption{Heatmap results of ResNet-50 (left), DenseNet-121 (middle) and DenResCov-19 (right). Red ellipses indicate the (human-generated) detection areas that are correctly identified by the network and black ellipses represent the undetected areas from the corresponding network. The successfully classified images are annotated with green ticks and the wrongly classified images are annotated with red crosses.}
 \label{cxh}
\end{figure*}

\subsection{Heatmap analysis}
\label{sec:heatmap}
From the results presented in Table \ref{tab3}, we can easily observe that our network has achieved improved results in all four different datasets. Figures~\ref{cxrw} and \ref{cxh} illustrate the heatmap  results of our pipeline and two state-of-the-art networks based on the images from DXR4 dataset.

\par Figure~\ref{cxrw} highlights the main steps of our pipeline. The CXR image initializes the network. The outputs of the four blocks of ResNet-50 and DenseNet-121 are then concatenated. This concatenation creates four new heatmaps $56$, $28$, $14$, and $7$ (gray squares). The new heatmaps $56$, $28$, and $14$ pass from a convolution and average max-pool block layer (Conv-blockA, Conv-blockB, and Conv-blockC), so that the dimensions of exported images can be decreased to $7 \times 7$ pixels. Following this step, the network combines the Conv-blockA with the Conv-blockB outputs (A-concat) and the Conv-blockC with the new-heatmap $7$ (B-concat). The last step is a concatenation of A-concat and B-concat to extract the Global-concat heatmap. Based on that, the model learns to classify the images in the supervised training tasks. Figure~\ref{cxrw} demonstrates the delineation of the COVID-19 detection points, denoted by red ellipses, by our proposed pipeline from a CXR lung image. The delineated region is identical to the area of interest detected by an expert radiologist.

Figure~\ref{cxh} shows the heatmaps of DenseNet-121, ResNet-50, and DenResCov-19 from a total of eight classification cases. In all cases of Fig.~\ref{cxh}, we have highlighted the last heatmap layer of the networks. The red circle in the CXR images are the detection points from our expert radiologist (AS). These points are used to classify the disease in each CXR image. In the top figure of Fig.~\ref{cxh}, all images are accurately diagnosed by the three networks. The black and red circles in the heatmap images denote the wrong and accurate detection points, respectively, with respect of the manual annotation. The extraction of the circle is based on a colormap threshold of $0.5$. If the average number of the area inside the circle is higher than the threshold, then the detection point assumes correct (red circle); otherwise, it assumes wrong (black circle). In the bottom figure of Fig.~\ref{cxh}, the DenseNet-121 accurately diagnoses the last two images (green tick), but wrongly classifies the first two (red cross). On the other hand, the ResNet-50 accurately diagnoses the first two images (green tick), while wrongly classifies the last two (red cross). In comparison, our network diagnoses correctly all images except the first one.

In the top figure of Fig.~\ref{cxh}, the DenseNet-121 cannot detect the left circle annotation (black circle) in the COVID-19 CXR image. Additionally, it cannot clearly detect the right circle annotation either. On the other hand, the ResNet-50 cannot detect the right circle annotation (black circle), while it identifies the left circle. In comparison, DenResCov-19 can detect both circle annotations strongly. In the tuberculosis CXR image, the DenResCov-19 also strongly detects the circle annotation, while the ResNet-50 and DenseNet-121 cannot detect clearly the circle annotation (near the threshold). However, in the pneumonia CXR image, the DenseNet-121 strongly detects the red circle annotation, while the ResNet-50 and DenResCov-19 cannot clearly detect it.

\subsection{Discussions}

One important limitation of this study is the relatively small cohort size for patients with COVID-19. Due to this, we mixed the pediatric and adult patients populations of the sources-1, source-2 and 3 in DXR4 dataset, in order to test the robustness of our proposed model in a dataset with larger cohort size. To avoid the bias effects, we created the dataset with randomly selected balanced number of images. Even if we test our pipeline in a huger dataset than DXR3 or DXR2; we had to deal with the limitation of mixed ages in the population (adults and children CXR images). As a result, there are some detection features, for example the pneumonia scars in Fig.~\ref{cxh} (top frame) or the healthy case in Fig.~\ref{cxh} (bottom frame) in DXR4 dataset, which our pipeline cannot strongly detect. The detection of these pathologies in DXR4 is more challenging and a larger dataset with additional demographics is required for further investigation.

Another limitation of this study is the multi-label lung pathology task. In order to further evaluate the generalization and robustness of our pipeline as a whole lung multi-pathologies classifier, we need to provide the multi-class and multi-label classification. Although we have delivered the multi-class challenge in the best possible way based on the available published cohorts, we still face a luck of the generalization from the multi-label aspect. An example of multi-label sample is when a subject has both bacterial pneumonia and COVID-19 diseases. The main reason we could not deliver this aspect is the lack of any publicly available multi-label lung disease datasets. 

\par The main advantage of this study is that, in the majority of lung pathologies, the detection points of radiologist CXR lung images are identified more strongly using the DenRes- Cov-19 network, as compared to the ResNet-50 and DensNet-121. The heatmap results presented in Fig.~\ref{cxrw} and \ref{cxh} justify the accurate classification of our network and validate our initial hypothesis. Moreover, all the evaluation metrics of the different classification datasets (from DXR1 to DXR4) demonstrate the robustness and superior performance of the DenResCov-19 network as compared to the benchmark deep learning based approaches.

\par As we discussed in Section~\ref{sec:related-work}, there are some limitations in the majority of the existing studies regarding robust and efficient detection of the Covid-19 and lung diseases. In our current study, we have examined these limitations and tried to solve them. We have trained our models based on regularization techniques, such as data augmentation and penalty L2 norms, to avoid possible overfitting. Furthermore, we have verified the generalization and accurate prediction of our model using Monte Carlo cross-validation techniques. The proposed method is fully automated and it does not need any manual segmentation of the lung region from experts to deliver a robust classification result. Finally, we have demonstrated different applications of the model over binary and multi-class classification tasks.

\section{Conclusions}
\label{sec:conclusions}

In this study, we have implemented a new deep-learning network named DenResCov-19, which can deliver robust classification results in multi-class lung diseases. We have tested the proposed model over three different published datasets with four classes, namely, the COVID-19 positive, pneumonia, TB, and healthy patients. We have also mitigated the class imbalance issue by properly composing the datasets (except for DXR4, where the dataset is imbalanced in COVID-19 positive class due to limited number of available images). Hence, based on our experimental analysis, we can infer a favorable generalization and robust behavior of our proposed model. Our experimental analysis has demonstrated improved classification accuracy of our network, as compared to the state-of-the-art networks such as ResNet-50, DenseNet-121, VGG-16, and Inception-V3. Our initial hypothesis that our network can deliver a well-balanced AUC-ROC and F1 metric results has been verified. In most of the cases, the detection points of our network from heatmaps are in line with the detection points from the expert radiologist. To summarize, we have developed a pre-screening fast-track decision network to detect COVID-19 and other lung pathologies based on CXR images.

\par In our future study, we will further focus on the generalization of our model with the availability of a significantly larger COVID-19 patients’ cohort. In addition, it will be beneficial to extend the number of classes to include more lung diseases if the corresponding datasets exist. Finally, we wish to evaluate the DenResCov-19 network in different datasets, in order to further evaluate the generalization and robustness of our pipeline in different medical image classification tasks, such as diagnosing multi-label lung diseases and other medical disease classification.

\section*{Acknowledgements}
The work of Andrew J. Swift was supported by the Wellcome Trust fellowship grant 205188/Z/16/Z. The authors acknowledge the use of facilities of the Research Software Engineering (RSE) Sheffield, UK. The authors express no conflict of interest.


\bibliographystyle{cas-model2-names.bst}
\bibliography{covid-19-cxr-ieee.bib}

\end{document}